\documentclass[dvips]{article}

\usepackage{icrc2011}

\newcommand{\be}{\begin{equation}}
\newcommand{\ee}{\end{equation}}
\newcommand{\ba}{\begin{eqnarray}}
\newcommand{\ea}{\end{eqnarray}}

\newcommand{\bfi}{\begin{figure}
\epsfxsize=9cm
\epsffile}
\newcommand{\efi}{\end{figure}}
\newcommand{\bi}{\begin{itemize}}
\newcommand{\ei}{\end{itemize}}

\title{Observation of the Cosmic Ray Moon shadowing effect with the ARGO-YBJ experiment}

\newcommand{\etal}{\MakeLowercase{\textit{et al. }}} 
\shorttitle{author \etal paper short title}

\authors{G. Di Sciascio$^{1}$ and R. Iuppa$^{1,2}$ for the ARGO-YBJ collaboration}
\afiliations{$^1$INFN, Sezione Roma Tor Vergata, Via della Ricerca Scientifica 1, Roma, Italy\\ $^2$ Dipartimento di Fisica, Universit\'a Roma Tor Vergata, via della Ricerca Scientifica 1, Roma, Italy}
\email{disciascio@roma2.infn.it, iuppa@roma2.infn.it}

\abstract{
Cosmic rays are hampered by the Moon and a deficit in its direction is expected (the so-called \emph{Moon shadow}). The Moon shadow is an important tool to determine the performance of an air shower array. In fact, the displacement of the shadow center, due to the bending effect of the Geomagnetic field on the propagation of cosmic rays, allows to set the energy scale of the primary particles inducing the showers observed by the detector. The shape of the shadow permits to determine the detector point spread function. The position of the deficit at high energy allows evaluating its pointing accuracy. Here we present the observation of the cosmic ray Moon shadowing effect carried out by the ARGO-YBJ experiment (Yangbajing Cosmic Ray Laboratory, Tibet, P.R. China, 4300 m a.s.l., 606 g/cm$^2$ ) in the multi-TeV energy region with high statistical significance (70 standard deviations). By means of an accurate Monte Carlo simulation of the cosmic rays propagation in the Earth-Moon system we have studied the role of the Geomagnetic field and of the detector point spread function on the observed shadow.}

\keywords{ Extensive Air Showers, Cosmic Rays, Moon shadow, Geomagnetic field, ARGO-YBJ}

\begin{document}
\maketitle

\section{Introduction}

The analysis of the Moon shadow observed by an air shower array may provide unique information on its performance.
At high energies, the Moon shadow would be observed by an ideal
detector as a 0.52$^{\circ}$ wide circular deficit of events,
centered on the Moon position. 
The actual shape of the deficit as reconstructed by the detector allows the determination of the angular resolution while the position of the deficit allows the evaluation of the absolute pointing accuracy.
In addition, charged particles are deflected by the Geomagnetic field (GMF) by an angle depending on the energy. As a consequence, the observation of the displacement of the Moon shadow at low rigidities can be used to determine the relation between the shower size and the primary energy.
The same shadowing effect can be observed in the direction of the
Sun but the interpretation of this phenomenology is less straightforward \cite{argo-solar}.
Finally, the Moon shadow can be exploited to measure the antiproton content in the primary cosmic rays (CR). In fact, acting the Earth-Moon system as a magnetic spectrometer, paths of primary antiprotons are deflected in the opposite sense with respect to those of the protons in their way to the Earth. This effect has been used to set limits on the $\bar{p}$ flux at TeV energies not yet accessible to balloon/satellite experiments \cite{antip}.

In this paper we present the observation of the cosmic ray Moon shadowing effect carried out by the ARGO-YBJ experiment during the period from July 2006 to November 2010.
We report on the angular resolution, the pointing accuracy and the energy scale calibration of the detector in the multi-TeV energy region. The results are compared with the predictions of a detailed simulation of cosmic ray propagation in the Earth-Moon system.

\section{The ARGO-YBJ experiment}

The detector is composed of a central carpet 74$\times$
78 m$^2$, made of a single layer of Resistive Plate Chambers
(RPCs) with about 93$\%$ of active area, enclosed by a guard ring
partially (20$\%$) instrumented up to $\sim$100$\times$110
m$^2$. The apparatus has a modular structure, the basic data
acquisition element being a cluster (5.7$\times$7.6 m$^2$),
made of 12 RPCs (2.85$\times$1.23 m$^2$ each). Each chamber is
read by 80 external strips of 6.75$\times$61.80 cm$^2$ (the spatial pixels),
logically organized in 10 independent pads of 55.6$\times$61.8
cm$^2$ which represent the time pixels of the detector. 
The read-out of 18360 pads and 146880 strips are the experimental output of the detector \cite{aielli06}.
The central carpet contains 130 clusters (hereafter, ARGO-130) and the
full detector is composed of 153 clusters for a total active
surface of $\sim$6700 m$^2$. 

The whole system, in smooth data taking since July 2006 firstly with ARGO-130, is in stable data taking with the full apparatus of 153 clusters since November 2007 with the trigger condition N$_{trig}$ = 20 and a duty cycle $\geq$85\%. The trigger rate is $\sim$3.5 kHz with a dead time of 4$\%$.

The analysis reported in this paper refers to events collected
after the following selections: (1) more than 20 strips on the ARGO-130 carpet; (2) zenith angle of the shower arrival direction less than 50$^{\circ}$; (3) reconstructed core position inside a 150$\times$150 m$^2$ area centered on the detector. According to the simulation, the median energy of the selected protons is E$_{50}\approx$1.8 TeV (mode energy $\approx$0.7 TeV).

\section{Monte Carlo simulation}

A detailed Monte Carlo (MC) simulation has been performed in order to propagate the CRs in the Earth-Moon system \cite{mcmoon-icrc09}.
The simulation is based on the real data acquisition time. The Moon position has been computed at fixed times, starting from July 2006 up to November 2010. Such instants are 30 seconds apart from each other. For each time, after checking that the data acquisition was really running and the Moon was in the field of view, primaries were generated with arrival directions sampled within the Moon disc according to the effective exposure time.

After accounting for the arrival direction correction due to the magnetic bending effect, the air shower development in the atmosphere has been generated by the CORSIKA v6.500 code with QGSJET/GHEISHA models \cite{corsika}. 
CR spectra have been simulated in the energy range from 10 GeV to 1 PeV following the compilation results given in \cite{horandel,wiebels}. About 10$^8$ showers have been sampled in the zenith angle interval 0-60 degrees. 
The experimental conditions (trigger logic, time resolution, electronic noises, etc.) have been reproduced by a GEANT4-based code \cite{geant4}. 

The trajectory of a primary cosmic ray (energy $E$, charge $Z$) crossing the GMF is observed by a detector placed at Yangbajing bent along the East-West (EW) direction, whereas no deviation is expected along the North-South (NS) one (see Fig. \ref{DataMCEW}, upper plot). To perform an evaluation of the bending effect, we adopted the Tsyganenko-IGRF model \cite{Tsyganenko}, which takes into account both internal and external magnetospheric sources by using data available from spacecraft missions.
To a first approximation, the amount of the EW shift can be written as: $\Delta\alpha\simeq-1.58^{\circ}\frac{Z}{E[\textrm{TeV}]}$.
The sign is set according to the usual way to represent the EW projection of the Moon maps. This equation can be easily derived by assuming that the GMF is due to a pure static dipole lying in the centre of the Earth and is valid for nearly vertical primaries with energy greater than a few TeV.
%
\begin{figure}[!t]
  \centering
  \includegraphics[width=0.42\textwidth]{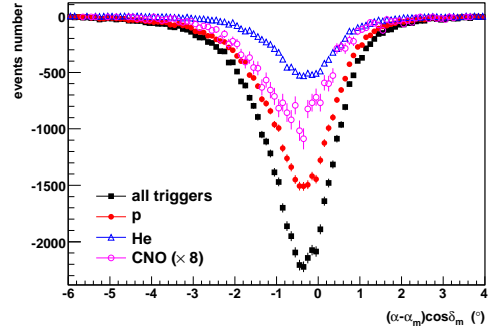}\\
  \caption{Simulated deficit counts around the Moon projected to the EW axis for N$_{strip}>$100. The contribution of different primaries to the Moon shadow deficit can be appreciated. The CNO component has been multiplied by a factor 8.
  \label{fig:components}}
   \end{figure}

The contribution of different cosmic ray primaries (protons, Helium and CNO group) to the Moon shadow deficit is shown in Fig. \ref{fig:components}. 
Events contained in an angular band parallel to the EW axis and centered on the observed Moon position, compatible with the multiplicity-dependent angular resolution, are used.

The angular width of the Moon contributes to the spread of the signal, therefore we must disentangle this effect in measuring the detector angular resolution.
Assuming a Gaussian PSF with variance $\sigma_{\theta}^2$, the width of the observed signal results:
\begin{eqnarray}
\label{eq:rmsmoon} RMS=\sigma_{\theta}\sqrt{1+\left(\frac{0.52^{\circ}}{4\cdot\sigma_{\theta}}\right)^2}
\end{eqnarray}
The contribution of the Moon size to the RMS is dominant when $\sigma_{\theta}$ is low, i.e. at high particle multiplicities. For instance, the difference between $RMS$ and $\sigma_{\theta}$ is 
$20\%$ if $\sigma_{\theta}$ = 0.2$^{\circ}$, less than $5\%$ if $\sigma_{\theta} >$0.4$^{\circ}$, and only $1.7\%$ if $\sigma_{\theta}$ = 0.7$^{\circ}$.
The effect of the detector angular resolution along the EW projection of the Moon shadow deficit determines not only the smearing, but also a further displacement of the signal peak due to the folding with the asymmetrical deflection induced by the GMF. The West tail of the shifted signal, indeed, has a larger weight compared with the sharp East edge and tends to pull the signal in its direction.

\section{Results}

\begin{figure}[!t]
  \centering
  \includegraphics[width=0.42\textwidth]{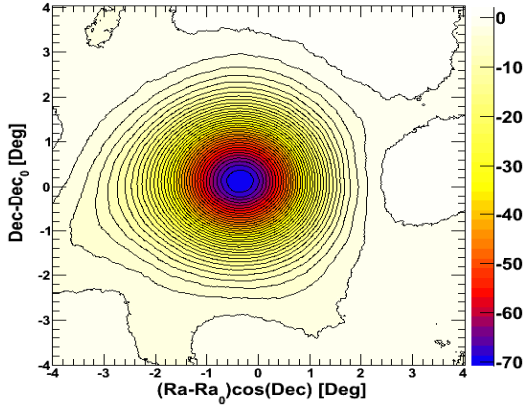}\\
 \caption{Significance map of the Moon region for events with N$_{strip}>$100, observed by the ARGO-YBJ experiment in the period July 2006 - November 2010. The coordinates are R.A. and DEC. centered on the Moon position.
The color scale gives the statistical significance in terms of standard deviations.
  \label{fig:moon-map}}
   \end{figure}

In Fig. \ref{fig:moon-map} the significance map of the Moon region observed with data recorded in the period July 2006 - November 2010 is shown for events with fired strips N$_{strip}>$ 100. The opening angle $\psi$ used in the smoothing procedure is 1$^{\circ}$. The statistical significance of the maximum deficit is about 70 standard deviations. The ARGO-YBJ experiment is observing the Moon shadow with a significance of about 9 standard deviations per month.
%
%

\subsection{Angular resolution}

%
\begin{figure}[!htpb]
  \centering
  \includegraphics[width=0.42\textwidth]{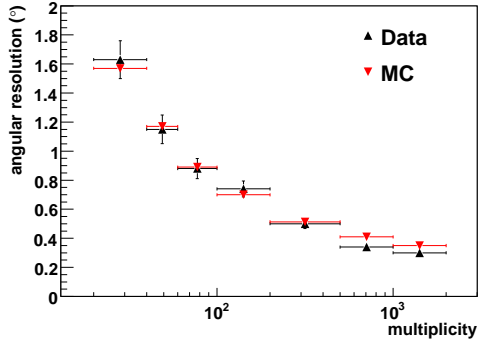}\\
  \caption{Measured angular resolution of the ARGO-YBJ detector (upward black triangles) compared to expectations from MC simulation (downward red triangles) as a function of the particle multiplicity. The multiplicity bins are shown by the horizontal bars. 
\label{fig:angresol}}
   \end{figure}
%
The PSF of the detector, studied in the NS projection not affected by the GMF, is Gaussian for N$_{strip}\geq$200, while for lower multiplicities is better described for both MC and data with an additional Gaussian, which contributes for about
20\%. For these events the angular resolution is calculated as the
weighted sum of the $\sigma_{\theta}^2$ of each Gaussian.
In Fig. \ref{fig:angresol} the angular resolution measured along the NS direction is compared to MC predictions as a function of the particle multiplicity, i.e. the number of fired strips N$_{strip}$ on ARGO-130. The values are in fair agreement showing that the ARGO-YBJ experiment is able to reconstruct events starting from only 20 particles spread on an area $\sim$6000 m$^2$ large with an angular resolution better than 1.6$^{\circ}$.

\subsection{Absolute energy scale calibration}

In order to calibrate the absolute energy scale of CRs observed by the ARGO-YBJ detector we can use the GMF as a magnetic spectrometer. 
In fact, the westward displacement of CRs by an angle inversely proportional to their energy provides a direct check of the relation between the shower size and the primary energy.
%
\begin{figure}[!htpb]
  \centering
  \includegraphics[width=0.42\textwidth]{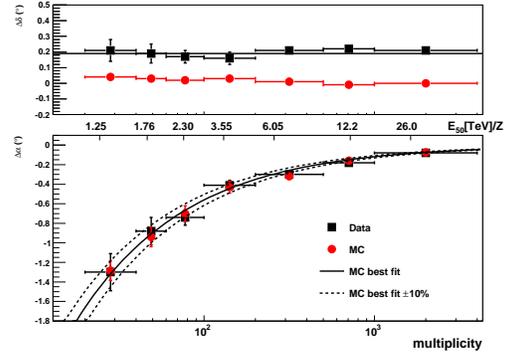}\\
\caption{Measured displacements of the Moon shadow (upper plot: NS, lower plot: EW) as a function of multiplicity (black squares). The data are compared to MC expectations (red circles). In the upper plot the solid line is fitted to the data.
In the lower plot the solid curve is fitted to the MC events and the dashed curves show the $\pm$10\% deviation from the solid one.
The energy scale refers to the rigidity (TeV/Z) associated to the median energy in each multiplicity bin (shown by the horizontal bars).
  \label{DataMCEW}}
   \end{figure}
%
In Fig. \ref{DataMCEW} the displacements of the Moon shadow in both NS (upper plot) and EW (lower plot) directions as a function of the particle multiplicity are shown.
The upper scale refers to the rigidity (TeV/Z) associated to the median energy in each multiplicity bin.
The observed shift is compared to the results of the MC simulation of CR propagation in the Earth-Moon system. 
A shift of (0.19$\pm$0.02)$^{\circ}$ towards North can be observed. This displacement is independent of the multiplicity.
Concerning the EW direction, the good agreement between data and simulation allows the attribution of this displacement to the combined effect of the detector PSF and the GMF. 
Therefore, the energy scale can be fixed in the multiplicity range 20-2000 particles, where the Moon shadow is moving under the bending effect of the GMF.
The MC results are fitted by the function 
$\Delta\alpha=\kappa (N_{strip})^{\lambda}$, with $\kappa$ = -10.17 and $\lambda$ = -0.63, shown by the solid curve in Fig. \ref{DataMCEW}, lower plot.
To estimate the possible shift in particle multiplicity between data and simulation, as shown by the dotted curves in Fig. \ref{DataMCEW}, the experimental data are fitted by the same function but with a multiplicity shift term: $-10.17[(1-\Delta R_{n}) N_{strip}]^{-0.63}$, as described in \cite{amenomori09}.
The parameter $\Delta R_{n}$ is the multiplicity shift ratio, resulting in $\Delta R_{n}$ = (+4 $\pm$ 7)$\%$.  
Finally, the conversion from $\Delta R_{n}$ to the energy shift ratio $\Delta R_{E}$ is performed. To determine the relationship between $\Delta R_{n}$ and
$\Delta R_{E}$, and to check that this method is sensitive to energy, six MC event samples in which the energy of the primary particles is systematically shifted event by event in the Moon shadow simulation are calculated \cite{amenomori09}. These six $\Delta R_{E}$ samples correspond to $\pm 20\%$, $\pm 15\%$ and $\pm 8\%$. Finally, by assuming a linear dependence, the relation  $\Delta R_{n}$= (-0.91 $\pm$ 0.16)$\cdot\Delta R_{E}$ is obtained.
Hence, the systematic uncertainty in the absolute energy scale $\Delta R_{E}$ is estimated to be (+5 $\pm$ 8)$\%$, where the error is the statistical one.

Two systematic uncertainties may affect this analysis, the first related to the assumed primary CR composition and the second to the use of different hadronic interaction models. We evaluated both contributions following \cite{amenomori09}: $\sigma_{chem}$ = 7$\%$, $\sigma_{hadr}$ = 12$\%$.

Finally, the absolute energy scale uncertainty in the ARGO-YBJ experiment is estimated to be smaller than $\sqrt{\Delta R_{E}^{2} + \sigma^{2}_{stat} + \sigma^{2}_{chem} + {(\sigma_{hadr}/2)}^{2}}$ = 13$\%$ in the energy range from 1 to 30 (TeV/Z), where the Moon shadow is shifted from its position due to the effect of the GMF.

\subsection{Long-term stability of the detector}

%
\begin{figure}[!htpb]
  \centering
  \includegraphics[width=0.42\textwidth]{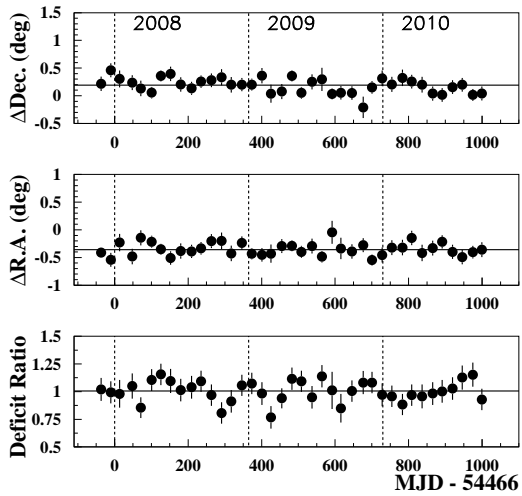}\\
  \caption{Upper panel: Displacement of the Moon shadow from the apparent center in the NS direction as a function of the observation time.
  Middle panel: Displacement of the Moon shadow from the apparent center in the EW direction as a function of the observation time.
  Lower panel: The ratio of the observed deficit count to the expected one as a function of the observation time. The plots refer to events with a multiplicity N$_{strip}>$100.
  \label{Data-stability}}
   \end{figure}
%
The stability of the detector performance as far as the pointing accuracy and 
the angular resolution are concerned is crucial in $\gamma$-ray astronomy.
Since November 2007 the full detector is in stable data taking with duty cycle $\geq$85\%.
Therefore, the stability of the ARGO-YBJ experiment has been checked by monitoring both the position of the Moon shadow, separately along R.A. and DEC. projections, and the amount of shadow deficit events in the period November 2007 - November 2010, for each sidereal month and for events with N$_{strip}>$100. 

The displacement of the center of the Moon shadow in the NS direction enables us to estimate the systematic error in pointing accuracy and its long-term stability aside from MC simulations, since the EW component of the GMF is almost zero at Yangbajing. The displacement of the shadow position from the Moon center in the NS direction is plotted in the upper panel of Fig. \ref{Data-stability} as a function of the observation time. 
Assuming a constant function, the best fit result (continuous line) shows that the Moon shadow is shifted towards North by (0.19$\pm$0.02)$^{\circ}$. The RMS around this position is 0.13$^{\circ}$.

In the middle plot the displacement along the EW direction is shown. The best fit result (continuous line) shows that the Moon shadow is shifted westward by (-0.36$\pm$0.02)$^{\circ}$, in agreement with the MC expectations ((-0.35$\pm$0.07)$^{\circ}$). The RMS around this Moon position is 0.11$^{\circ}$.

The amount of CR deficit due to the Moon provides a good estimation of the size of the shadow, therefore of the angular resolution. 
The expected number of deficit events N$_{def}$ is approximatively given by
$N_{def}=\left[1-e^{-0.5\cdot(\frac{R}{\sigma})^2}\right]\cdot N_{moon}$,
where R is the angular distance from the source, N$_{moon}$ is the number of events intercepted by the Moon and $\sigma$ is the Gaussian width of the shadow.
In the lower plot of Fig. \ref{Data-stability} the ratio of the observed 
deficit count to the expected one is shown as a function of the observation time. The expected deficit events have been calculated for $\sigma$=0.8$^{\circ}$. 
The line shows the best fit result assuming a constant function: 1.005$\pm$0.016. The RMS of the corresponding distribution is 0.11.

\section{Conclusions}

The Moon shadowing effect on cosmic rays has been observed by
the ARGO-YBJ experiment in the multi-TeV energy region with a statistical significance greater than 70 standard deviations.

We have estimated the primary energy of the detected showers by measuring the westward displacement as a function of the multiplicity, thus calibrating the relation between shower size and CR energy.
The systematic uncertainty in the absolute energy scale is evaluated to be less than 13\% in the range from 1 to 30 (TeV/Z). 

The position of the Moon shadow measured with the ARGO-YBJ experiment turned out to be stable at a level of 0.1$^{\circ}$ and the angular resolution stable at a level of 10\%, on a monthly basis.
These results make us confident about the detector stability in the long-term observation of gamma-ray sources.



\begin{thebibliography}{99}

\bibitem{argo-solar} G. Aielli et al., \emph{ApJ} \textbf{729}, 113 (2011).
\bibitem{antip} G. Di Sciascio and R. Iuppa, these proceedings. \bibitem{aielli06} G. Aielli et al., \emph{NIM} \textbf{A562}, 92 (2006).
\bibitem{mcmoon-icrc09} G. Di Sciascio and R. Iuppa, \emph{NIM} \textbf{A630}, 301 (2011).
\bibitem{corsika} D. Heck et al., Report \textbf{FZKA 6019} (1998).
\bibitem{horandel} J.R. Horandel, \emph{Astrop. Phys} \textbf{19}, 193, 2003.
\bibitem{wiebels} B. Wiebel-Sooth et al, \emph{A\&A} \textbf{330}, 389, 1998.
\bibitem{geant4} Y.Q. Guo et al.,\emph{Chinese Phys.} \textbf{C34} (2010) 555.
\bibitem{Tsyganenko} N.A. Tsyganenko, \emph{J. Geophys. Res.} \textbf{100}, 5599 (1995). 
\bibitem{amenomori09} M. Amenomori et al., \emph{ApJ} \textbf{692}, 61 (2009).

\end{thebibliography}
\end{document}